\magnification=1200

\centerline{\bf DEFLAGRATION TO DETONATION}

\bigskip\centerline{A.M. Khokhlov}
\bigskip\centerline{Laboratory for Computational Physics and Fluid Dynamics,}
        \centerline{Naval Research Laboratory, Washington, DC}

\input epsf.sty

\newcount\notenumber
\notenumber=0
\def\note{\global\advance\notenumber by 1 
	  \footnote{$^{\the\notenumber}$}}

\def\msun{{M$_\odot$}~}

\bigskip\noindent
 Thermonuclear explosions of Type Ia
supernovae (SNIa) involve turbulent deflagrations, detonations,
and possibly a deflagration-to-detonation transition. 
A phenomenological delayed
detonation model of SNIa successfully explains many observational properties of
SNIa including monochromatic light curves, spectra,
brightness -- decline and color -- decline relations.
Observed variations among SNia are explained as a result of varying
nickel mass synthesised in an explosion of a Chandrasekhar mass C/O white dwarf.
Based on theoretical models of SNIa, the value of the Hubble constant 
$H_o\simeq 67$km/s/Mpc was determined without the use of secondary distance
indicators. 
The cause for the nickel mass variations in SNIa is still debated. It may be a 
variation of the initial C/O ratio in a supernova progenitor, rotation,
or other effects.

\bigskip
\bigskip\centerline{\bf 1. Introduction}
\bigskip
Type Ia supernovae (SNIa) are important astrophysical objects which are
increasingly used as distance indicators in cosmology.
SNIa appear to be a rather well behaving group of objects. 
There are deviations in maximum
brightness of $\sim 2$m among SNIa, but they
correlate with variations in the shape of SNIa light curves:
less bright supernovae tend to decline faster.
This is often expressed as a correlation between
$m$ and $dm_{15}$, where $dm_{15}$ is the decrease in magnitude
15 days since maximum. Another
correlation exists between SNIa color at maximum and postmaximum
decline, $(B-V)$ -- $dm_{15}$ -- less bright supernovae tend to be more red [1]. 
These two correlations can be used to account for variations 
in brightness of SNIa and for interstellar absorption. Using these 
has led to improved determinations of $H_o$ [2] and to new findings 
concerning 
$\Omega_m$ and $\Omega_\Lambda$ [3].

Are there exact and unique
 maximum brightness~--~postmaximum decline and color~--~postmaximum decline 
relations among SNIa? Are these relations the same for nearby and 
cosmological supernovae? 
Before these questions so important for $\Omega_m$ and $\Omega_\Lambda$ can
be addressed from theoretical grounds,
we would like  the theory of SNIa to answer more
general questions: 

\itemitem{(1)} Why do SNIa differ from each other?
\itemitem{(2)} Why do some of SNIa characteristics correlate?

\bigskip\centerline{\bf 2. Pre-supernovae}

\bigskip
It is believed that SNIa are thermonuclear explosions of 
carbon-oxygen (CO) white dwarfs (WD).
However, evolutionary paths leading to SNIa are 
still a bit of a mystery [4]. 
Three major scenarios have been 
considered based on the evolution of binary stellar systems: (1) a CO-WD 
accreting
mass through Roche-lobe overflow from an evolved companion  star
[5]. The explosion is triggered by compressional heating near the WD
center when the WD approaches the Chandrasekhar mass.
(2)  Merging of  two low-mass WDs caused by the loss of angular momentum due 
to gravitational radiation [6]. Resulting merged configuration 
consists of a massive WD component surrounded by the
rotationally supported envelope made of 
less massive, disrupted WD [7]. 
If ignition takes place at low densities,
near the base of the rotating envelope, it will probably lead to slow
burning and subsequent core collapse [8].
Otherwise, gradual redistribution of angular momentum may lead to
a growth of a massive CO core which then ignites near the center 
when its mass approaches the Chandrasekhar limit.
The exploding configuration will resemble an isolated $\sim 1.4$\msun CO-WD,
but with rotation and surrounded by an extended CO envelope.
(3) a CO-WD accreting mass through Roche-lobe overflow as in (1), but
the explosion is triggered by the detonation of an accumulated layer of
helium
before the total mass of the configuration reaches the Chandrasekhar mass
[9].  Only the first two models appear to be viable. The third, the 
sub-Chandrasekhar WD model, has been ruled out on the basis of
predicted light curves and spectra. 

\bigskip\centerline{\bf 3. Phenomenological models of SNIa explosion}

\bigskip
Many ingredients of the SNIa explosion physics such as equation of
state and nuclear reaction rates are known well. 
However, flame propagation in a supernova is difficult
to model from first principles due to an enormous disparity
 of spatial and temporal scales involved. One is forced to make 
assumptions about regimes of burning (detonation or deflagration), 
and about the speed with which the flame propagates in case of turbulent 
deflagration. Once the assumptions are made, an outcome 
can be calculated by solving the equations of fluid dynamics
coupled with the (prescribed) nuclear energy release terms, and with terms describing
self-gravity of the star. 

Three major models of the
explosion of a Chandrasekhar mass CO-WD have been considered: (1) detonation 
model 
[10], (2) deflagration model 
[11],
and (3) delayed-detonation (DD) model [12-15]. 
The most detailed computations of SNIa explosion to date involve a hydrodynamic
calculation of the thermonuclear explosion that includes a nucleosynthesis
computation,  a time-dependent radiation-transport computation that gives the
light curve, including mechanisms for $\gamma$-ray and positron deposition, the
effects of expansion opacity, and scattering,  and NLTE spectra computations
[16,17 and references therein].

It was found that purely detonation models do not fit observations
 because they do
not produce intermediate mass (Si-group) elements which are so prominent in
the spectra of SNIa around maximum light.
Deflagration models produce intermediate mass elements, but typically in a too
narrow velocity range, and also have difficulty explaining the variety of
SNIa. Delayed detonation models
are successful in reproducing the main features of  SNIa,
including multi-wavelength light curves, the spectral behavior, and
the brightness -- decline and color -- decline correlations [16,18].

Delayed detonation models assume that burning starts as a subsonic
deflagration and then turns into a supersonic detonation. The deflagration
speed,  $S_{\rm def}$,  and the moment of deflagration-to-detonation
transition (DDT) 
 are free parameters.
The moment of DDT is conveniently parametrized by introducing the transition
density, $\rho_{\rm tr}$, at which DDT happens. 
Initial  central density 
 and initial  composition (C/O ratio) of the the exploding WD must also be
specified. 
To reproduce observations,
deflagration speed should be a rather small
fraction of the speed of sound $a_s$, say,  $S_{\rm def} < 0.1 a_s$.
Physical arguments why $S_{\rm def}$ is small are discussed in the 
next section.
The models are very sensitive to the variations of $\rho_{\rm tr}$, but   
to a much lesser extent on the exact assumed value of the deflagration speed,
initial central density of the exploding star, and the initial chemical
composition.

A delayed detonation explosion is schematically illustrated in Figure 1.
Because the speed of deflagration is less than
the speed of sound, pressure waves generated by burning propagate ahead
of the deflagration front and cause the star to expand.
As a result, deflagration  propagates through
matter which density continuously decreases with time. 
After deflagration turns into a detonation, detonation wave 
incinerates the rest of the WD left unburned during the deflagration phase. 
Detonation produces Fe-group elements if it occurs at 
densities greater than $\rho \simeq 10^7$g/cc.
At lower densities it produces intermediate mass elements.
At even lower densities around $\sim 10^6$g/cc only carbon has time to burn.
The outermost layers of a supernova 
will consist of products of explosive carbon burning such as O, Ne, Mg, etc.
To reproduce observations,  $\rho_{\rm tr}$ must be selected in the
 range  $\rho_{\rm tr} \simeq (1-3)\times 10^7$g/cc.
Virtually no intermediate mass elements will be produced for larger values of
$\rho_{\rm tr}$.
For lower $\rho_{\rm tr}$, the WD expands
 so much that a detonation cannot be sustained.
With  $\rho_{\rm tr}$ in the right range,  the inner parts of the exploded  
star consist
of  Fe-peak elements and contain radioactive  $^{56}$Ni. Outer parts contain
intermediate group elements and products of explosive carbon burning (Figure 1).

The amount of $^{56}$Ni produced during the
explosion  is very sensitive to $\rho_{\rm tr}$. Varying $\rho_{\rm tr}$ in 
the range 
$(1-3)\times 10^7$g/cc gives nickel mass in the range $\simeq 0.1 - 0.7
M_\odot$, respectively. The reason for such a sensitivity is the combination
of an exponential 
temperature dependence of  reaction rates and the
dependence of the specific heat of a degenerate matter on density. Small 
differences in
density at which burning takes place translate into small differences in
burning temperature.  These, however, translate into large
differences in reaction rates, and into qualitative differences in the resulting
chemical composition. 
The kinetic energy of the explosion, on the other hand, is very insensitive
to  $\rho_{\rm tr}$. It depends on the total amount of burned material 
(Fe-group and Si-group together).
This is because the difference in binding energies of Fe-group and
Si-group nuclei is relatively small compared to the difference between
binding energies of both Fe- and Si-group
elements and the initial CO mixture. Thus, the delayed detonation model predict
SNIa  with significantly varying nickel mass but with
 almost the same kinetic energy and expansion velocities.

The above property of the delayed detonation model is 
the key to the explanation of the brightness -- decline and  
color -- decline relations among Type Ia supernovae.
All delayed detonation supernovae expand with approximately the same velocity.
Explosions with more nickel give rise to brighter supernovae. 
Also, because of more nickel decays, envelopes of these supernovae are heated
better and stay hot and opaque. 
The result is a slow post-maximum decline and a blue color. Explosions with
less nickel give rise to dim supernovae. Envelopes of these supernovae are
cool and transparent because they contain less nickel.
The result is a fast post-maximum decline and a red color\note{In 
deflagration 
models, the amount of nickel and 
kinetic energy of the explosion are tightly related. Supernovae with
more nickel expand  and cool faster, while supernovae with less nickel are 
expanding slowly. Deflagration models predict that light curves of
brighter supernovae should decline faster, which is contrary to
observations.}.

As a representative example, Figure 2 shows results of numerical modeling of
the bright SNIa 1994D [19]. 
The light curves of SN1994D are fit with the  light
curves of the best fit delayed detonation model M36,  one of the models of the
series of [16] with the initial central density 
$\rho_c =2.7\times 10^9$ g/cc and $S_{\rm def} = 0.04 a_s$. For the M36 model, the
transition density was $\rho_{\rm tr} = 2.4\times 10^7$ g/cc.   As can be seen,
both optical and IR light curves are fit by M36 rather well, including the
secondary maximum in R and I typical of normally bright SNIa. Models with other
values of $\rho_{\rm tr}$ led to much worse fits to observations. 

Delayed detonation models have been used to predict  a purely theoretical 
(without using Cepheid distances) value of the Hubble constant 
[20,16].
The idea is to fit a supernova with the model which best reproduces its
light curves and spectra. The model then gives the absolute brightness of
a supernova. The method takes into account both brightness 
variations among SNIa and possible interstellar absorption. The result is shown 
in Figure~3. Values of $H_o$ determined from individual supernovae show a large
spread for close SNIa but converge to $H_o\simeq 67\pm 9$km/s/Mpc with
increasing $z$. This value is in agreement with 
$H_o$ found using Cepheid variables [21].

\bigskip\centerline{\bf 4. Three-dimensional SNIa}

\bigskip
Three-dimensional  effects 
in the propagation of turbulent flames and 
deflagration-to-detonation transition (DDT) must
 play a key role in SNIa in determining
the actual speed of the  flame propagation, energetics, and nucleosynthesis,
and also are likely to  translate initial differences in presupernova structure
into the observed  differences among SNIa. 

\medskip
{\sl Deflagration.} -- Laminar flame in a WD is driven by heat conduction
due to degenerate electrons and propagates very subsonically with the speed
 $S_{\rm lam} <0.01 a_s$ [22]. Such a slow flame cannot  
account for the explosion properties of SNIa. However, in the presence of
gravity, the flame speed will be enhanced  by the Rayleigh-Taylor instability
[23]. Whether the Rayleigh-Taylor instability can itself sufficiently 
increase the flame speed  to cause the explosion or whether deflagration
just  serves to  pre-expand the star which is incinerated later by a
supersonic  detonation, has been a subject of numerous studies
and 3D simulations[24,25,27,32 and references therein]. 

Simple scaling arguments show that turbulent flame  
subjected to a uniform gravity acceleration in a vertical column must
 propagate with a speed [24]
$$
S_{\rm def} \simeq 
       \alpha \left(gL{\rho_0-\rho_1\over\rho_0 + \rho_1}\right)^{1/2}~,\eqno(1)
$$
 where
 $g$ is  local acceleration, $L$ is the width of the column, 
 $\rho_0$ and $\rho_1$ are densities ahead and behind the flame,
respectively, and $\alpha < 1$ is a constant which depends on the
column's geometry and boundary conditions. Formula (1) is valid, of
course, only when $S_{\rm def}$ is much larger than $S_{\rm lam}$. It tells
that when the characteristic RT speed $\simeq (gL)^{1/2}$ is greater than
 $S_{\rm lam}$, the flame speed is determined by the turbulence on
the largest scale, $L$, independent of details of flame propagation on smaller
scales. The reason for this behavior is self-similarity of the flame. 
Turbulent flame speed is the product of the area $A$ of the flame surface
and the laminar flame speed, $S_{\rm def} = A \cdot S_{\rm lam}$.
Turbulence tends to increase $A$ whereas intersections
of different portions of the flame front tend to decrease $A$.
The latter effect is proportional to $S_{\rm lam}$. In equilibrium, the two
effects balance each other, and $A\propto 1/S_{\rm lam}$. 
The product of $A$ and $S_{\rm lam}$ remains 
constant\note{There is a close analogy here with the
self-similarity of an ordinary Kolmogorov cascade. In the Kolmogorov
 cascade, changes in the fluid
viscosity lead to changes in the viscous microscale, but do not influence
the amount of energy dissipated into heat. The rate of dissipation depends
only on the intensity of turbulent motions on the largest scale. $S_{\rm
 lam}$ plays the role of viscosity in a turbulent flame. }.
Three-dimensional numerical simulations with varying $g$, $L$
and varying laminar flame speed confirm equation (1) including its
independence of $S_{\rm lam}$ and indicate $\alpha\simeq 0.5$ [24,26].
The results are consistent with high-gravity combustion experiments that
used a   centrifuge to study premixed
turbulent flames at various $g$.

Equation (1) tells several important things. Near the WD center $g\simeq
0$. Thus, in the beginning the deflagration speed $S_{\rm def} \simeq S_{\rm
lam}$ should be small.
The speed will then tend to increase as the
flame goes away from the center and the gravity increases.
When the intensity of turbulence increases, the flame speed will become
independent of the physics of burning on small scales. The latter conclusion
is very important since it gives us a hope that SNIa explosions can be modeled
in three dimensions without resolving all small spatial and temporal scales.
However, equation (1) is missing an important piece of physics. It is valid
only in a uniform gravitational field and only when there is no global
expansion of matter. 

In a supernova explosion, burning causes a global
expansion of a star. Equation (1) may be valid only on scales where
the expansion velocity is less than the characteristic RT speed. On larger
scales, expansion will tend to freeze the turbulence out. The net result
will be a substantial decrease of the turbulent flame speed. 
A  crude estimate of the scale $L_f$ at which the turbulence becomes frozen
 and of the effective deflagration speed limited by expansion 
can be obtained as follows. First, carry out a one-dimensional
 simulation assuming no turbulence freeze-out, that is, with the flame speed 
given by equation (1) with $L$ equal 
to the flame radius $R_f$. This gives the expansion rate.
Then estimate $L_f$ as a scale at which the 
expansion velocity becomes comparable with the characteristic RT velocity.
Finally, estimate the effective
 deflagration speed from equation (1) using $L=L_f$ [27].
The estimates are $L_f\simeq ({\rm a~few})\times 10^7$cm, and 
$$
S_{\rm def} \simeq 1.5\times 10^7{\rm cm/s} \,
                    \left(g\over 10^9{\rm cm/s^2}\right)^{1/2}
                    \left(L_f\over 10^8 {\rm cm} \right)^{1/2}~.
                                                                     \eqno(2)
$$
Equation (2) shows that in conditions typical of
the exploding white dwarf, a turbulent burning speed
is a few per cent of the sound speed $a_s \simeq 5\times 10^8$cm/s.  
This is not enough to cause a powerful explosion.
An additional effect that further limits the rate of deflagration 
is a deviation from the steady-state turbulent burning regime. 
A certain time is  required for a turbulent
 flame to reach a steady-state. This time is
larger for larger scales. Scales of the order of $R_f$ might never reach a
steady-state during the explosion.

Figure 4 shows some results of a three-dimensional numerical simulation of
the entire  Chandrasekhar mass CO-WD  exploding as a supernova.
In this simulation, equation
(1) has been used  for the turbulent flame
velocity on scales not resolved numerically. For regions of the ``average''
flame front not oriented ``upwards'' against gravity this formula 
most probably overestimates the local turbulent flame speed. 
Despite this,  only $\simeq 5$\% of the mass has been burned by
the time the star has expanded and quenched the flame, and
the white dwarf has not even become unbound. These results show that spherical
expansion is indeed important and that burning
on large scales does not reach a steady state.  Big blobs of burned
gas rise and penetrate low density outer layers, whereas unburned
matter flows down and reaches the stellar center. 
The model experienced
an almost complete overturn. This has obviously important implications
for nucleosynthesis and may cause an element stratification 
incompatible with observations if composition inhomogeneities are not 
smeared out during the subsequent detonation stage of burning.
The results of 3D modeling indicate that the deflagration
alone is not sufficient to cause an explosion.  To make a powerful explosion, 
the deflagration must somehow make a 
transition to a detonation (delayed detonation
model). 

{\sl Deflagration-to-Detonation Transition.} -- 
In terrestrial conditions detonation
may arize from a non-uniform
explosion of a region of a fuel with a gradient of reaction (induction) time
via the Zeldovich gradient mechanism [28]. The region may be
created by mixing of fresh fuel and hot products of burning, as in 
 jet initiation, or it may be created by multiple
shocks, etc. The same gradient mechanism can operate in supernovae
[29-31].
There exist a minimum, critical size of the  region capable of 
generating a detonation, $L_i$. This parameter is determined by the 
the equation of state and nuclear reaction rates and is mainly a function of
the density of the material.  $L_i$ is much much less than
the size of a WD for all but very low densities $\rho< 10^7$g/cc [30]. 

Why then DDT does not happen in supernovae at high
densities? Why does it have to wait until the WD expands significantly? The 
explanation may be this. The critical size $L_i$, however small, is still several
orders of magnitude larger than the thickness of a laminar flame. To mix 
fresh fuel with products of burning, the surface of the flame must be
disrupted. But this is difficult to achieve unless the
 turbulence on a scale of a flame front is larger than the laminar flame
speed. Only at very low densities, where reactions slow down, the width of the
laminar flame becomes very large, and its speed becomes very small,
 the turbulence may have a chance  
to create the right conditions for DDT [30,31]. 

In addition to mixing fuel and products inside an active deflagration front,
another mechanism for creating the right conditions for DDT may be as follows.
 As mentioned above, turbulence in a SNIa will be limited by the expansion.
The conditions for DDT during the expansion of a star may not be fulfilled
at all. But when deflagration speed is small, deflagration quenches due to expansion
before  the WD becomes unbound. This happens, in particular,
 in the simulation shown in Figure 4. The star will then experience a
pulsation  and collapse back. During the expansion and contraction phases 
of the pulsation, the high entropy ashes of dead deflagration front will mix
with the fresh low entropy fuel again to form a mixture with the reaction time 
gradients. Mixing will be facilitated during the
contraction phase due to the increase of turbulent motions due to
the conservation of angular momentum (like a skating ring dancer increase his
rotation by squizing his arms). The estimate of the mixing region formed
during pulsation is $\simeq 10^6 - 10^7$ cm, much larger than $L_i$. Is was
also shown that as soon as only a few per cent of hot ashes are mixed with a cold
fuel, the mixture cannot be compressed to densities higher than
$\simeq ({\rm a~few}) \times 10^7$g/cc. Further compression will lead to
a burnout on time scales much shorter than the pulsation  time scale.
 As soon as this mixture returns to high enough densities
$\simeq 10^7$ and re-ignites, the detonation will be triggered
[13,33,30].

It should be noted at this point that
three-dimensional theory of flame propagation and DDT in supernovae is far
from being finished, and remains a subject of an active research.
In particular, it was speculated recently that DDT may be caused 
by a sudden acceleration of a quasi-spherical deflagration front, due 
to the Landau-Darrheus or some other yet unknown internal 
instabilities of the flame; that a suddenly accelerated deflagration might
keep propagating with the speed of sound without turning into a detonation, etc. [32].
Whether any of these can actually happen should be either tested in
appropriately scaled terrestrial experiments or demonstrated in
three-dimensional simulations. Further work is required, and it 
will undoubtedly improve our understanding of SNIa explosions.

It may also be possible to to distinguish between different multi-dimensional
explosion mechanisms on the basis of observations.
One of the amazing properties of SNIa is their apparently small 
deviation from spherical symmetry. We do not expect all three-dimensional
models to have this property. For example, pure deflagration models
are expected to be clumpy (Figure 4) and asymmetric with large blobs of Si and
Fe group elements embedded in the unburned CO envelope.
Delayed detonation models, on the other hand, should be more
symmetric. A supersonic detonation mode of burning that
follows deflagration will tend to homogenize the ejecta. 
 Rotation of the progenitor may
impose a global, low order asymmetry on the ejecta. 
Viable models can be limited by computing the polarization of the emerging
radiation and comparing the predictions with the existing [34]
and planned observations. 

\bigskip\centerline{\bf 5. Discussion}

\bigskip
We described a phenomenological delayed detonation model of
SNIa based on the explosion of a Chandrasekhar mass
carbon-oxygen white dwarf. The model assumes that the explosion starts
as a subsonic deflagration and then turns into a supersonic detonation mode
of burning. The model is successful in reproducing the main features of  SNIa,
including multi-wavelength light curves, the spectral behavior, and
the brightness -- decline and color -- decline correlations.
It was  argued that an apparently low deviations of SNIa from spherical
symmetry (low polarization of SNIa) may be attributed in delayed detonation
models to the homogenizing effect of the detonation phase of an explosion.

The model interprets existing brightness -- decline
and color -- decline relations among SNIa as a result of varying 
nickel mass synthesised during the explosion.
Major free parameters of the model are the deflagration speed
$S_{\rm def}$, the transition density  $\rho_{\rm tr}$ at which deflagration
turns into a detonation, and also initial density and composition (C/O
ratio)  of the exploding WD. The variation of nickel mass in the model
is caused by the variation of $\rho_{\rm tr}$.
Strong sensitivity of the nickel mass to $\rho_{\rm tr}$ is probably the 
basis of why, 
to first approximation, SNIa appear to be a one-parameter family. 
Nonetheless, variations of the other parameters also lead to some relatively
small variations of the predicted properties of SNIa, which indicate that 
the assumption of a one-parameter family may not be strictly valid.

To fit observations, the delayed detonation model requires low values
of $S_{\rm def} < 0.1 a_s$ and low values 
of $\rho_{\rm tr} \simeq (1-3)\times 10^7$g/cc. 
In Section 4 it was argued that slow deflagration is the result
of an expansion of a star caused by the deflagration itself. The expansion 
tends to freeze the turbulence and, thus, limits the deflagration speed.
The actual rate of deflagration in a supernova is determined by the
competition of the Rayleigh-Taylor instability which is the turbulence
driving force, the turbulent cascade from large to small scales, and the 
turbulence freeze-out. 
Two possible mechanisms that lead to a low  $\rho_{\rm tr}$ were discussed --
one related to the disruption of an active deflagration front by the
existing turbulence, and the other related to quenching of deflagration,
mixing of the low-entropy fuel with high-entropy burning products,
 and its subsequent compression. 

It may seem unusual that the two
apparently different mechanisms predict almost the same low values for
$\rho_{\rm tr}$,  these same low values that are required to fit
observations
in
phenomenological delayed detonation models.
Note, however, that predictions of low transition density by both DDT
 mechanisms and the very reason why 
low $\rho_{\rm tr}$ is needed to fit SNIa observations steam from the same two 
fundamental facts: (1) specific heat of matter in supernovae depends on
density;  (2) nuclear reactions depend on temperature exponentially. 
The resulting dependence of burning timescales on density is very
steep. 
Numbers are such that at densities above $10^7$g/cc nuclear burning
timescales are much shorter
than  the sound crossing time ($\simeq $ explosion timescale) of a WD.
At densities below $10^7$g/cc the timescales become much longer than the sound
crossing time. 
That is why a laminar flame front can 
 be disrupted by turbulence only below approximately $10^7$g/cc -- flame
width is
proportional to a burning timescale and at higher densities it is
 much much shorter than any other relevant spatial scale of a WD.
That is why a mixture of cold fuel and hot products cannot be 
compressed to densities much higher than $ 10^7$g/cc --
at higher densities it will react faster than it is being compressed. 
And that is also why intermediate mass elements can be synthesised in an
SNIa  only at densities around $\sim 10^7$g/cc -- at higher densities
reactions will
have enough time to reach a nuclear statistical equilibrium
 and, thus, to burn CO into Ni-group elements.

What may cause the variations of $\rho_{\rm tr}$ among SNIa?
There are several possibilities. One is differences in the initial C/O
ratio. 
If less
carbon is present near the WD center, less energy will be released by burning, and this will
affect both the buoyancy of burning products and the rate of expansion of a
WD.  This, in turn, will affect the speed of deflagration, and will
lead to different conditions for DDT. Variations of initial C/O ratio
among SNIa has been recently studied in the framework of one-dimensional
phenomenological delayed detonation models in [35-37]. The effect of varying
C/O ratio may result in small but noticeable variations in the rise time to
maximum
and in some other variations in light curve behavior. This is a potential
source of systematic evolutionary effects, and has obvious
implications for using SNIa in cosmology.
However, in one-dimensional models one has to assume how changes in C/O
influence the deflagration speed and  $\rho_{\rm tr}$, and the predictions
then depend on these assumptions. Three-dimensional
modeling is required in order to predict the actual influence of C/O ratio
on the outcome of the explosion.

Another possibility is the influence of rotation if SNIa are the result of a
merger of low-mass CO-WD. Rotation will undoubtedly influence the turbulent
deflagration phase which, in turn, will affect DDT. Merger configurations
may also differ in their mass, so that slightly super-Chandrasekhar mass WD
explosions are probable. Could they be responsible for unusually bright
SN1991T-like events? An extended CO envelope around a merger WD may manifest
itself in SNIa light curves and spectra. 
Further work is needed to answer these questions.

\medskip\noindent
This contribition is based in part on work done 
 in collaboration
with Peter H\"oflich, Ewald
M\"uller,  Elaine Oran, Craig Wheeler, and others. I thank them and also 
David Arnett, David Branch, Caren Brown, Robert Harkness, Eli Livne, 
Ken Nomoto, Geraint Thomas, and Lifan Wang for many discussions.  
This research was supported in part by the NASA Grant NAG52888
and by the Office of Naval Research.

\bigskip

\centerline{\bf References}
\bigskip
\hang\noindent
1. Phillips, M. M. 1993, ApJ, 413, L105 

\hang\noindent
2. Riess A.G., Press W.H., Kirshner R.P. {1996}, {ApJ}, {473}, {588} 

\hang\noindent
3. Shmidt, B. et al., 1998, ApJ, 507, 46; Perlmutter, S. et al. 1999, ApJ, 517,565.

\hang\noindent
4. Livio, M. 2000, in The Largest Explosions Since the Big Bang: Supernovae
   and Gamma-Ray Bursts, eds. M. Livio, K. Sahu, \& N. Panagia, in press.

\hang\noindent
5. Nomoto, K. \& Sugimoto, D., 1977, PASJ, 29, 765; Nomoto, K., 1982, ApJ
   253, 798.

\hang\noindent
6.  Webbink  R.F. 1984, ApJ, 277, 355; 
    Iben  I.Jr., Tutukov  A.V. 1984, ApJS, 54, 335.

\hang\noindent
7. Benz, W., Bowers, R.L., Cameron, A.G.W, Press, W.H., 1990, ApJ, 348, 647.

\hang\noindent
8. Mochkovich R., Livio, M. 1990, A\&A, 236,378.

\hang\noindent
9. Nomoto  K. 1980, ApJ ,248, 798; Woosley S.E., Weaver T.A., Taam
R.E. 1980,  in: Type I Supernovae, eds. C.Wheeler, Austin, U.Texas, p. 96.

\hang\noindent
10. Arnett  W.D. 1969, Ap. Space Sci., 5, 280; Hansen  C.J., Wheeler
J.C. 1969,  Ap. Space Sci., 3, 464.

\hang\noindent
11. Nomoto  K., Sugimoto  S., \& Neo  S. 1976, ApSS, 39, L37.

\hang\noindent
12. Khokhlov A.M. 1991, AA, 245, 114.

\hang\noindent
13. Khokhlov A.M. 1991, AA, 245, L25.

\hang\noindent
14. Yamaoka H., Nomoto K., Shigeyama T., Thielemann F.-K. 1992, ApJ, 393, 55.

\hang\noindent
15.     Woosley, S. E., Weaver, T. A. 1994, in Les Houches, Session LIV,
Supernovae,  ed. S. A. Bludman, R. Mochkovich, \& . Zinn-Justin (Amsterdam:
   North-Holland), 63

\hang\noindent
16. H\"oflich, P., Khokhlov, A.M.  1996, ApJ, 457, 500

\hang\noindent
17. Nugent, P. et al.  1997, ApJ, 485, 812

\hang\noindent
18.       H\"oflich, P.; Khokhlov, A.; Wheeler, J. C.  1995, ApJ, 444, 831;
     H\"oflich, P.; Khokhlov, A.,  Wheeler, C.J., Phillips, M.M., 
     Sunzeff, N.B., Hamuy, M. 1996, ApJ, 472, L81;
     Wheeler et al. 1998, ApJ, 496, 908, and references therein.

\hang\noindent
19. H\"oflich, P. 1995, ApJ, 459, 307.

\hang\noindent
20. M\"uller, E., H\"oflich, P.A.  1994, A\&A, 281, 51.

\hang\noindent
21. Sandage, A. 2000, in The Largest Explosions Since the Big Bang: Supernovae
   and Gamma-Ray Bursts, eds. M. Livio, K. Sahu, \& N. Panagia, in press.; 
    Riess, A., Nugent, P., Filippenko, A.V., Kirshner, R., Perlmutter, S.,
1998, ApJ, 504, 935; Ferrarese, L. et al. 1999, ApJ, in press.

\hang\noindent
22. Timmes F.X., Woosley S.E. 1992, ApJ 396, 649.

\hang\noindent
23. Nomoto  K., Sugimoto  S., \& Neo  S. 1976, ApSS, 39, L37

\hang\noindent
24. Khokhlov, A.M., 1995, ApJ, 449, 695.

\hang\noindent
25.  Livne, E., Arnett, W.D. 1993, ApJ 415,
L107; Arnett, W.D., Livne, E. 1994, 427, 314.; 

\hang\noindent
26. Khokhlov, A.M. Oran, E.S., Wheeler, J.C. 1996, Combustion \& Flame,
    105, 28

\hang\noindent
27. Khokhlov, A.M., Oran, E.S., Wheeler, J.C., 1995,  in Type Ia Supernovae,
       proc. of the NATO conference, Barcelona, Spain, 1995.

\hang\noindent
28. Zeldovich, Ya. B., Librovich, V. B., Makhviladze, G. M., \& Sivashinsky,
    G. L. 1970, Acta Astron., 15, 313;
    Lee, J. H. S., Knystautas, R., \& Yoshikawa, N. 1978, Acta Astron., 5,
    971.

\hang\noindent
29. Blinnikov, S.I., Khokhlov, A.M. 1986, Soviet Astron. Lett., 12, 131.

\hang\noindent
30. Khokhlov, A.M., Oran, E.S., Wheeler, J.C. 1997, ApJ, 478, 678.

\hang\noindent
31. Niemeyer, J.C., Woosley, S.E. 1997, ApJ, 475, 740. 

\hang\noindent
32.  Niemeyer, J.C., 1999, ApJ, 523, L57.

\hang\noindent
33.  Arnett, D., \& Livne, E. 1994, ApJ, 427, 330.

\hang\noindent
34. Wang, L., Wheeler, J.C., H\"oflich, P. 1997, ApJ, 476, 27.

\hang\noindent
35. Dominguez, I., H\"oflich, P.A., 1999, ApJ, in press (astro-ph/9908204);
    H\"oflich, P.A., Dominguez, I., 2000, in The Largest Explosions Since
    the Big Bang:
    Supernovae
   and Gamma-Ray Bursts, eds. M. Livio, K. Sahu, \& N. Panagia, in press.

\hang\noindent
36. Umeda, H., Nomoto, K., Kobayashi, C., Hachisu, I., Kato, M., 1999, ApJL,
    in press

\hang\noindent
37. H\"oflich, P.A., Nomoto, K., Umeda, H., Wheeler, J.C., 1999, ApJ, in press.

\vfill\eject

\centerline{\bf Figure captions}

\bigskip\noindent
1. Schematics of the delayed detonation explosion. Ignition takes
place in a dense, Chandrasekhar mass carbon-oxygen white
dwarf. Flame propagates from the center as a subsonic turbulent
deflagration.
Deflagration-to-detonation transition (DDT) takes place in a significantly 
expanded star after only a small fraction of mass has been burned.
Detonation  incinerates the rest of the white dwarf. The resulting
configuration consists of the inner core of Fe-group elements including
$^{56}$Ni surrounded by a massive envelope of Si-group elements. 

\bigskip\noindent
2.  Comparison of observed (SN1994D) and theoretical (M36) B, V, R, I light 
    curves [19].

\bigskip\noindent
3. Direct determination of the Hubble constant ($H_o=67\pm 9$km/s/Mpc)
 using delayed detonation models [16].
Values of $H_o$ 
are plotted for individual SNIa based on distances determined by fitting 
their light curves and spectra with theoretical models.

\bigskip\noindent
4. Three-dimensional simulation of an explosion of a 
           Chandrasekhar mass CO-WD [24]. The figure shows density
           distribution during the deflagration phase of the explosion.

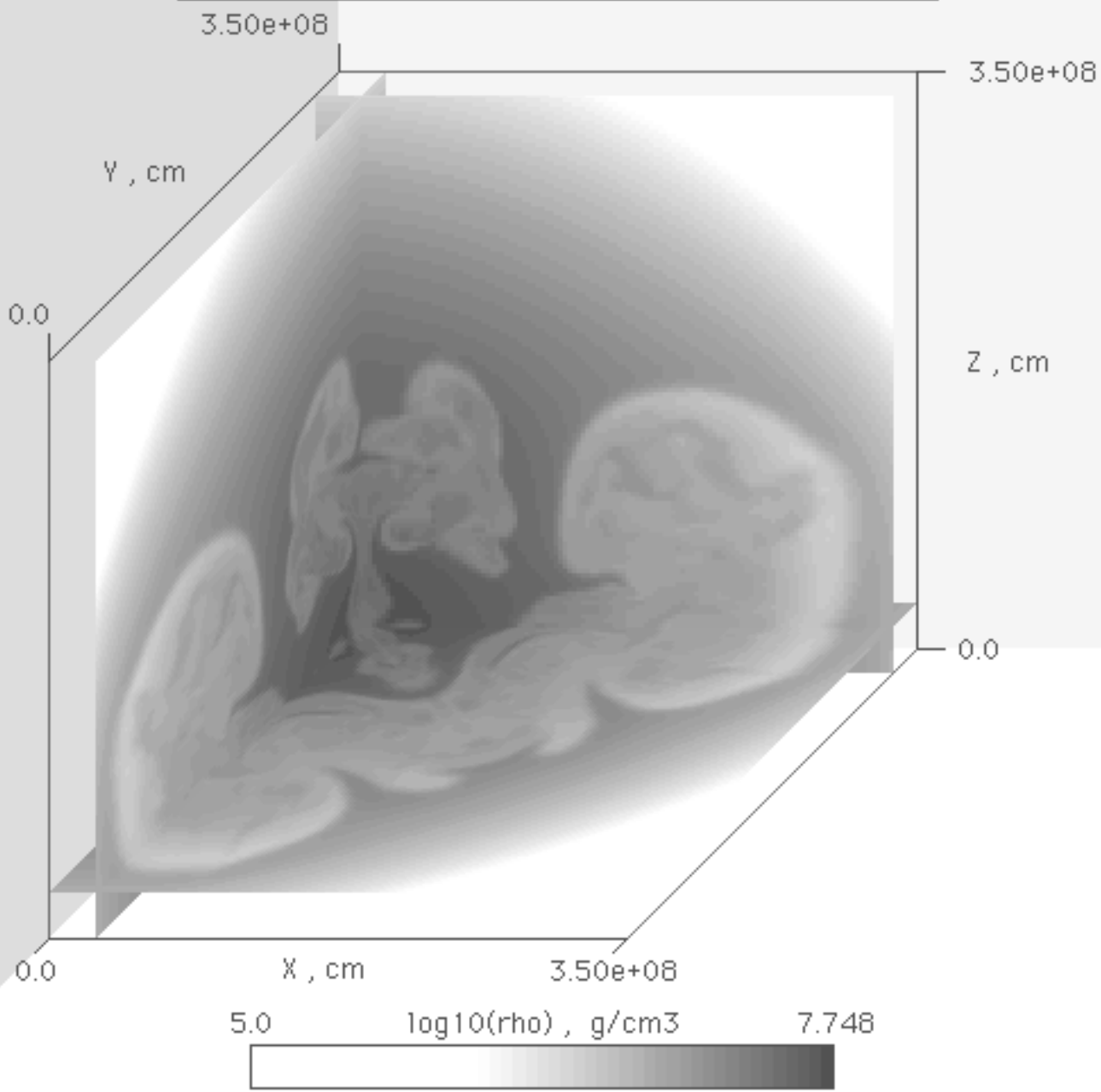
\end